\documentclass{PoS-hep}
\usepackage{amsmath}
\usepackage{bm}
\usepackage{epsfig}
\usepackage{graphics}
\usepackage{upgreek}
\usepackage{amsfonts}
\usepackage{amsbsy}
\usepackage{amscd}
\usepackage{bbm}
\usepackage{eufrak}
\usepackage{cite}

\renewenvironment{subequations}{%
\refstepcounter{equation}%
\setcounter{parentequation}{\value{equation}}%
  \setcounter{equation}{0}
  \ignorespaces
}{%
  \setcounter{equation}{\value{parentequation}}%
  \ignorespacesafterend
}

\newcommand{\Eref}[1]{Eq.~(\ref{#1})}
\newcommand{\Sref}[1]{Sec.~\ref{#1}}
\newcommand{\Fref}[1]{Fig.~\ref{#1}}

\newcommand{\cref}[1]{Ref.~\cite{#1}}

\newcommand{\bal}{\begin{align}}
\newcommand{\eal}{\end{align}}
\newcommand{\beqs}{\begin{subequations}}
\newcommand{\eeqs}{\end{subequations}}
\newcommand{\eec}{\end{center}}
\newcommand{\bec}{\begin{center}}

\newcommand{\eem}{\end{matrix}}
\newcommand{\bem}{\begin{matrix}}
\newcommand{\eeq}{\end{equation}}
\newcommand{\beq}{\begin{equation}}
\newcommand{\ba}{\begin{array}}
\newcommand{\ea}{\end{array}}
\newcommand{\bea}{\begin{eqnarray}}
\newcommand{\eea}{\end{eqnarray}}
\newcommand{\baq}{\begin{eqnarray}}
\newcommand{\eaq}{\end{eqnarray}}

\newcommand{\ftn}{\footnotesize}

\newcommand{\ssz}{\scriptsize}

\newcommand{\GeV}{{\mbox{\rm GeV}}}

\newcommand{\sFref}[2]{Fig.~\ref{#1}-{\ftn\sf ({#2})}}

\def\to{\rightarrow}

\def\lf{\left(}
\def\rg{\right)}
\newcommand\vev[1]{\langle {#1} \rangle}

\newcommand{\Vhi}{\ensuremath{\widehat V_{\rm HI}}}

\newcommand{\Hhi}{\ensuremath{\widehat H_{\rm HI}}}

\newcommand{\Khi}{\ensuremath{K}}

\newcommand{\Ns}{\ensuremath{{\what N_\star}}}

\newcommand{\mP}{\ensuremath{m_{\rm P}}}

\def\openone{\leavevmode\hbox{\small1\kern-3.8pt\normalsize1}}
\newcommand{\dV}{\ensuremath{\Delta\widehat V_{\rm HI}}}

\newcommand{\fr}{\ensuremath{f_{ R}}}

\newcommand{\phc}{\ensuremath{\Phi}}
\newcommand{\phcb}{\ensuremath{\bar\Phi}}

\newcommand{\ks}{\ensuremath{k_\star}}
\newcommand{\ns}{\ensuremath{n_{\rm s}}}

\newcommand{\as}{\ensuremath{a_{\rm s}}}
\newcommand{\As}{\ensuremath{A_{\rm s}}}
\newcommand{\rw}{\ensuremath{r_{0.002}}}
\newcommand{\rs}{\ensuremath{r_{\pm}}}

\newcommand{\rce}{\ensuremath{\widehat{{R}}}}
\newcommand{\Ve}{\ensuremath{\widehat{V}}}

\newcommand{\Dex}{\ensuremath{\Delta_{\rm max\star}}}

\newcommand{\what}{\ensuremath{\widehat}}

\def\bbet{{\bar\beta}}
\def\al{{\alpha}}

\def\bt{{\beta}}

\def\n{\bar{n}}

\def\thb{{\bar\theta}}

\def\thb{{\bar\theta}}
\def\thn{{\theta_{\Phi}}}

\newcommand{\sg}{\ensuremath{\phi}}

\newcommand{\sgx}{\ensuremath{\phi_\star}}
\newcommand{\sgf}{\ensuremath{\phi_{\rm f}}}

\newcommand{\ld}{\ensuremath{\lambda}}
\newcommand{\ldu}{\ensuremath{\uplambda}}

\newcommand{\kp}{\ensuremath{\kappa}}

\newcommand{\se}{\ensuremath{\widehat \phi}}

\newcommand{\geu}{\ensuremath{\widehat g}}

\def\Ka{K\"{a}hler potential}

\def\bcp{{\sc\small Bicep2}/{\it Keck Array}}
\newcommand{\plk}{{\it Planck}}

\newcommand{\cm}{\ensuremath{c_{-}}}
\newcommand{\cp}{\ensuremath{c_{+}}}
\newcommand{\fm}{\ensuremath{F_{-}}}
\newcommand{\fp}{\ensuremath{F_{+}}}

\newcommand{\rpm}{\ensuremath{r_{\pm}}}

\title{Observable Gravitational Waves from Higgs Inflation in SUGRA}

\ShortTitle{Observable Gravitational Waves from HI in SUGRA}

\author{\speaker{C. Pallis}\\
Department of Physics, University of Cyprus, \\ P.O. Box 20537,
Nicosia 1678, CYPRUS}

\abstract{We consider models of chaotic inflation driven by the
real parts of a conjugate pair of Higgs superfields involved in
the spontaneous breaking of a grand unification symmetry at a
scale assuming its Supersymmetric value. Employing quadratic \Ka s
with a prominent shift-symmetric part proportional to $\cm$ and a
tiny violation, proportional to $\cp$, included in a logarithm we
show that the inflationary observables provide an excellent match
to the recent \plk\ and \bcp\ results setting, e.g.,
$0.012\lesssim\cp/\cm\lesssim1/N$ where $-N<0$ is the prefactor of
the logarithm. Moreover, we analyze several possible stabilization
mechanisms for the non-inflaton accompanying superfield using just
quadratic terms. In all cases, inflation can be attained for
subplanckian inflaton values with the corresponding effective
theories retaining the perturbative unitarity up to the Planck
scale.
\\ \\ {\sl\bfseries
Published in}~~{PoS EPS-HEP {\bf 2017}, 047 (2017)}. }

\FullConference{The European Physical Society Conference on High Energy Physics\\
                  5-12 July 2017\\
                   Venice, Italy}
\begin{document}

\section{Introduction}

We focus on the simplest and most promising models of kinetically
modified non-minimal \emph{Higgs inflation} ({\sf\ftn HI})
established in \cref{nMHkin,var}. Namely, working in the context
of \emph{Supergravity} ({\sf\ftn SUGRA}), we concentrate on the
subclass of these models which employ a prominent non-trivial
kinetic coupling and only quadratic terms in the adopted \Ka s.
Moreover, we present novel stabilization functions of the
non-inflaton field inspired by \cref{su11}. We below describe the
formulation of this type of HI in the context of SUGRA -- see
\Sref{sugra} -- and then, in \Sref{inf}, we analyze the
inflationary behavior of these models. Our results are exposed in
\Sref{res} and our conclusions in \Sref{con}.

Throughout the text, the subscript $,\chi$ denotes derivation
\emph{with respect to} ({\sf\ftn w.r.t}) the field $\chi$, charge
conjugation is denoted by a star ($^*$) and we use units where the
reduced Planck scale $\mP = 2.43\cdot 10^{18}~\GeV$ is set equal
to unity.

\section{Modeling Higgs Inflation in SUGRA}\label{sugra}

In \Sref{sugra1} we present the basic formulation of a scalar
theory within SUGRA and then we outline in \Sref{sugra3} our
strategy in constructing viable models of HI.

\subsection{The General Set-up} \label{sugra1}

Our starting point is the \emph{Einstein frame} ({\sf\small EF})
action for the scalar fields $z^\al$ within SUGRA \cite{var} which
can be written as
\beqs \beq\label{Saction1} {\sf S}=\int d^4x \sqrt{-\what{
\mathfrak{g}}}\lf-\frac{1}{2}\rce +K_{\al\bbet}\geu^{\mu\nu} D_\mu
z^\al D_\nu z^{*\bbet}-\Ve\rg, \eeq
where $\rce$ is the Ricci scalar and $\mathfrak{g}$ is the
determinant of the background Friedmann-Robertson-Walker metric,
$g^{\mu\nu}$ with signature $(+,-,-,-)$. We adopt also the
following notation
\beq \label{Kab} K_{\al\bbet}={\Khi_{,z^\al
z^{*\bbet}}}>0\>\>\>\mbox{and}\>\>\>D_\mu z^\al=\partial_\mu
z^\al+ig A^{\rm a}_\mu T^{\rm a}_{\al\bt} z^\bt\eeq
are the covariant derivatives for scalar fields $z^\al$. Here, and
henceforth, the scalar components of the various superfields are
denoted by the same superfield symbol. Also, $g$ is the unified
gauge coupling constant, $A^{\rm a}_\mu$ are the vector gauge
fields and $T_{\rm a}$ are the generators of the gauge
transformations of $z^\al$. The EF potential, $\Ve$, is given in
terms of the K\"ahler potential, $K$, and the superpotential, $W$,
by
\beq \Ve=\Ve_{\rm F}+ \Ve_{\rm D}\>\>\>\mbox{with}\>\>\> \Ve_{\rm
F}=e^{\Khi}\left(K^{\al\bbet}{\rm F}_\al {\rm F}^*_\bbet-3{\vert
W\vert^2}\right) \>\>\>\mbox{and}\>\>\>\Ve_{\rm D}= {1\over2}g^2
\sum_{\rm a} {\rm D}_{\rm a} {\rm D}_{\rm a}. \label{Vsugra} \eeq
Here, the summation is applied over the generators $T_{\rm a}$ of
a considered gauge group -- a trivial gauge kinetic function is
adopted. Also we use the shorthand
\beq \label{Kinv} K^{\al\bbet}K_{\al\bar
\gamma}=\delta^\bbet_{\bar \gamma},\>\>{\rm F}_\al=W_{,z^\al}
+K_{,z^\al}W~~\mbox{and}~~{\rm D}_{\rm a}=z^\al\lf T_{\rm
a}\rg^\bt_\al K_\bt~~\mbox{with}~~
K_{\al}={\Khi_{,z^\al}}.\eeq\eeqs
In this talk we concentrate on HI driven by $\Ve_{\rm F}$ along a
D-flat direction, and therefore the contribution from $\Ve_{\rm
D}$ vanishes.

\subsection{Inflating With a Superheavy Higgs}\label{sugra3}

The general ideas above can be applied to HI if we employ three
chiral superfields, a conjugate pair, $z^1=\Phi$ and
$z^2=\bar\Phi$, {charged} under a local symmetry, e.g.
$U(1)_{B-L}$, and a gauge singlet $z^3=S$ which play the role of
``stabilizer'' superfield. We below present the utilized $W$
(\Sref{Wsec}) and $K$'s (\Sref{Ksec}).

\subsubsection{Superpotential}\label{Wsec}

\begin{floatingtable}[hr]
\begin{tabular}{|l||lll|}\hline
{\sc Superfields}&$S$&$\Phi$&$\bar\Phi$\\\hline\hline
$U(1)_{B-L}$&$0$&$1$&$-1$\\\hline
$R$ &$1$&$0$&$0$\\\hline
\end{tabular}
\caption {\sl Charge assignments of the superfields.}\label{ch}
\end{floatingtable}

Our scenario is based on the following superpotential
\beq W=\ld S\lf\bar\Phi\Phi-M^2/4\rg \label{W} \eeq
which is uniquely determined at renormalization level using a
$U(1)_{B-L}$ and an $R$ symmetry shown in Table~1. $W$ leads to a
$B-L$ phase transition at the scale $M$, which may assume the
value predicted by the SUSY unification -- see \cref{var} --,
since the SUSY vacuum lies at the direction
\beq \label{vevs} \vev{S}=0\>\>\>\mbox{and}\>\>\>
|\vev{\Phi}|=|\vev{\bar\Phi}|=M/2,\eeq
%
and so, $U(1)_{B-L}$ is spontaneously broken. Indeed, the SUSY
limit of $\Ve$, after HI, reads \beq V_{\rm
SUSY}=\ld^2\left|\phcb\phc-{M^2}/4\right|^2
+\frac{1}{\cm(1-N\rpm)}\ld^2|S|^2\lf |\phc|^2+|\phcb|^2\rg+{\rm
D-terms},\eeq where $N,\cm$ and $\rpm$ are defined below, and it
is minimized along the configuration of \Eref{vevs}.

\subsubsection{Possible K\"{a}hler Potentials}\label{Ksec}

\renewcommand{\arraystretch}{1.2}
\begin{table}[t]
\bec
\begin{tabular}{|c|c|c|}\hline
$F_{iS}$&{\sc Exponential Form} & {\sc Logarithmic
Form}\\\hline\hline

$F_{1S}$ & $\exp\lf-|S|^2/N\rg-1$ & $-\ln(1+|S|^2/N)$ \\
$F_{2S}$ & $-N_S\lf \exp\lf-|S|^2/N_S\rg-1\rg$ & $N_S\ln(1+|S|^2/N_S)$ \\
$F_{3S}$ & $-N_S\lf \exp\lf-\lf\cm\fm+|S|^2\rg/N_S\rg-1\rg$& $N_S\ln(1+\cm\fm/N_S+|S|^2/N_S)$ \\
\hline
\end{tabular}\label{tab0}\eec
\caption{\sl \small Functional forms of $F_{iS}$ with $i=1,2,3$
shown in the definition of $K_1, K_2$ and $K_3$ -- $N,N_S>0$.}
\end{table}
\renewcommand{\arraystretch}{1}

The proposed $W$ above may support HI if we combine it with one of
the following $K$'s
\beqs\bea \label{K1}&&
K_1=-N\ln\left(1+\cp\fp+F_{1S}(|S|^2)\right)+\cm\fm\,, \\
\label{K2} && K_2=-N\ln\left(1+\cp\fp\right)+\cm\fm +\
F_{2S}(|S|^2)\,,
\\ \label{K3}  && K_3=-N\ln\left(1+\cp\fp\right) +\
F_{3S}(\fm,|S|^2)\,,\eea\eeqs
where the functions $F_\pm=\left|\Phi\pm\bar\Phi^*\right|^2$
assist us in the introduction of shift symmetry for the Higgs
fields  -- cf. \cref{jhep} -- and the functions $F_{iS}$, given in
Table~2, assure the successful stabilization of $S$ along the
inflationary path. From the listed $F_{iS}$ only the logarithmic
forms for $i=2$ and $3$ are used until now in \cref{var}. In all
$K$'s, $\fp$ is included in the argument of a logarithm with
coefficient $-N$ whereas $\fm$ is outside it. The models can be
characterized as completely natural, because, in the limits
$\cp\to0$ and $\ld\to0$, they enjoy the following enhanced
symmetries:
\beq \bar\Phi \to\ \bar\Phi+c^*,\> \Phi \to\ \Phi+c \>\>(c \in
\mathbb{C})\>\>\>\mbox{and}\>\>\> S \to\ e^{i\alpha} S\,.\eeq


\section{Inflation Analysis}\label{inf}

We derive the tree-level inflatioanry potential in \Sref{inf1} and
then, in \Sref{inf2}, we check its robustness against corrections.

\subsection{Inflationary Potential}\label{inf1}

To study accurately enough the inflationary dynamics we use the
parametrization
\beq \Phi=\phi e^{i\theta}\cos\theta_\Phi/{\sqrt{2}}\>\>\>
\mbox{and}\>\>\>\bar\Phi=\phi e^{i\thb}\sin\theta_\Phi/{\sqrt{2}}
\>\>\>\mbox{and}\>\>\>S=({s +i\bar s})/{\sqrt{2}}\label{para}\eeq
with $0\leq\thn\leq{\pi}/{2}$. Then, we can show that a D-flat
direction is
\beq\theta=\thb=s=\bar
s=0\>\>\>\mbox{and}\>\>\>\thn={\pi/4}.\label{infltr}\eeq
Along it, the only surviving term of $\Ve_{\rm F}$ for any $K$ in
Eqs.~(\ref{K1}) -- (\ref{K3}) is
\beq \label{Vhi} \Vhi= e^{K}K^{SS^*}\,
|W_{,S}|^2=\frac{\ld^2(\sg^2-M^2)^2}{16\fr^{2(1+n)}}\>\>\>\mbox{where}\>\>\>\fr=1+\cp\sg^2\eeq
plays the role of a non-minimal coupling to gravity. Also, we set
\beq \label{ndef} n= \left\{\bem
%
(N-3)/2\hfill\cr
N/2-1\hfill\cr \eem
\right. \>\>\>\mbox{and}\>\>\> K^{SS^*}=\left\{\bem
%
\fr\hfill\cr
1\hfill\cr \eem
\right. \>\>\>\mbox{for}\>\>\> \left\{\bem
%
K=K_1\hfill\cr
K=K_{2} ~~\mbox{or}~~ K_3\,.\hfill\cr \eem
\right.\eeq

Note that, for $n>0$, $\Vhi$ develops a local maximum
\beq \Vhi(\sg_{\rm max})={\ld^2 n^{2 n}(1 + n)^{-2 (1 + n)}/{16
\cp^2}}~~\mbox{at}~~ \sg_{\rm max}=1/{\sqrt{\cp n}}\,.\eeq
Consequently, a tuning of the initial conditions is required which
can be quantified somehow defining the quantity
$\Dex=\left(\sg_{\rm max} - \sgx\right)/\sg_{\rm max}$, where
$\sgx$ is the value of $\sg$ when the pivot scale $\ks=0.05/{\rm
Mpc}$ crosses outside the inflationary horizon.

The EF canonically normalized fields, which are denoted by hat,
can be obtained as follows
\beq \label{can} \frac{d\se}{d\sg}=J,\>\>\widehat{\theta}_+
=\frac{J\sg\theta_+}{\sqrt{2}},\>\>\widehat{\theta}_-
=\sqrt{\frac{\kp_-}{2}}\sg\theta_-,\>\>\>\widehat \theta_\Phi =
\sg\sqrt{\kp_-}\lf\theta_\Phi-\frac{\pi}{4}\rg\>\>\>\mbox{and}\>\>\>
(\what s,\what{\bar{s}})=\sqrt{K_{SS^*}}(s,\bar s)\,,\eeq
where $J=\sqrt{\kp_+}$ with $\kp_+
=\cm\lf1+N\rpm(\cp\sg^2-1)/{\fr}\rg\simeq\cm$ and $\kp_-=\cm\lf1-
{N\rpm}/{\fr}\rg$. Also,
$\theta_\pm=(\theta\pm\bar\theta)/\sqrt{2}$. Positivity of $\kp_-$
requires ${\rpm<1/N}$ discriminating a little, thereby, the
domains of the solutions with $K=K_1$ and $K=K_2$ or $K_3$. Note
that $\cm$ influences only $J$ (and not $\Vhi$).

\subsection{Stability and Radiative Corrections}\label{inf2}

\renewcommand{\arraystretch}{1.2}
\begin{table*}[t]
\bec
\begin{tabular}{|c|c|c|c|c|c|}\hline
{\sc Fields}&{\sc Einge-} & \multicolumn{4}{c|}{\sc Masses
Squared}\\\cline{3-6}
&{\sc states} & & {$K=K_1$}&{$K=K_2$} &{$K=K_{3}$} \\
\hline\hline
2 Real &$\widehat\theta_{+}$&$\widehat m_{\theta+}^2$
&\multicolumn{2}{|c|}{$6\Hhi^2$}&$6(1+1/N_S)\Hhi^2$ \\\cline{3-6}
Scalars&$\widehat \theta_\Phi$ &$\widehat m_{ \theta_\Phi}^2$&
\multicolumn{2}{|c|}{$M^2_{BL}+6\Hhi^2$}&$M^2_{BL}+6(1+1/N_S)\Hhi^2$
\\\cline{3-6}
1 Complex Scalar&$\widehat s, \widehat{\bar{s}}$ & $ \widehat m_{
s}^2$&$6\cp\sg^2\Hhi^2/N$&\multicolumn{2}{c|}{$6\Hhi^2/N_S$}\\\hline
1 Gauge Boson & $A_{BL}$ &  $
M_{BL}^2$&\multicolumn{3}{c|}{$g^2\cm\lf1-N\rpm
/\fr\rg\sg^2$}\\\hline
$4$ Weyl & $\what \psi_\pm $ & $\what m^2_{ \psi\pm}$
&\multicolumn{3}{c|}{${6(\cp(N-
\mathfrak{n})\sg^2-2)^2\Hhi^2}/{\cm\sg^2\fr^{2}}$}\\\cline{2-6}
Spinors &$\ldu_{BL}, \widehat\psi_{\Phi-}$&
$M_{BL}^2$&\multicolumn{3}{c|}{$g^2\cm\lf1-N\rpm
/\fr\rg\sg^2$}\\
\hline
\end{tabular}\label{tab1}\eec
\caption{\sl \small  Mass-squared spectrum for $K=K_1, K_2$ and
$K_3$ along the path in Eq.~(2.2) taking
$\mathfrak{n}=3~[\mathfrak{n}=2]$ for $K=K_1$ [$K=K_2$ or $K_3]$.}
\end{table*}

\begin{figure}[!t]\vspace*{-.12in}
\hspace*{-.19in}
\begin{minipage}{8in}
\epsfig{file=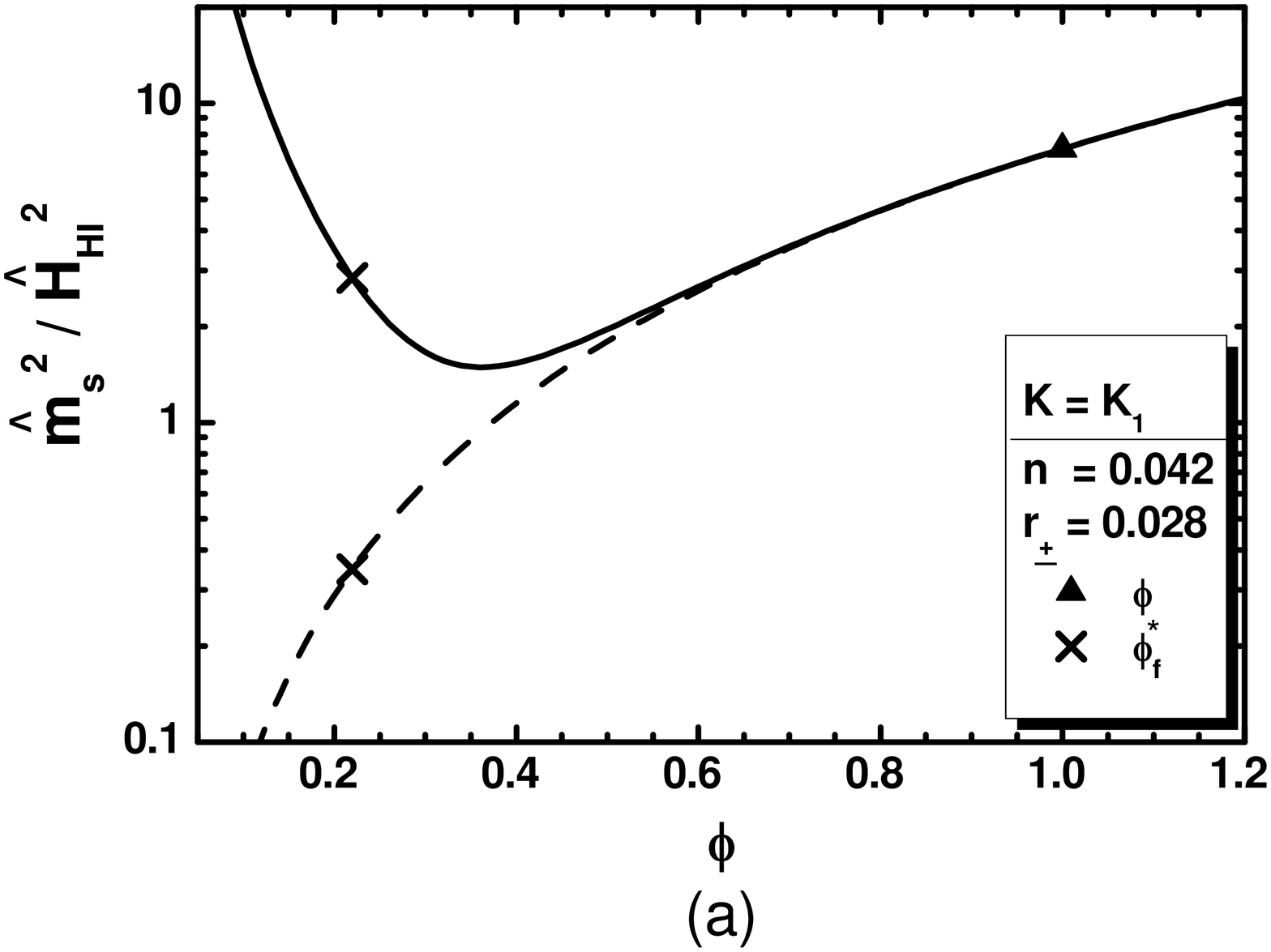,height=3.48in,angle=-90}
\hspace*{-1.2cm}
\epsfig{file=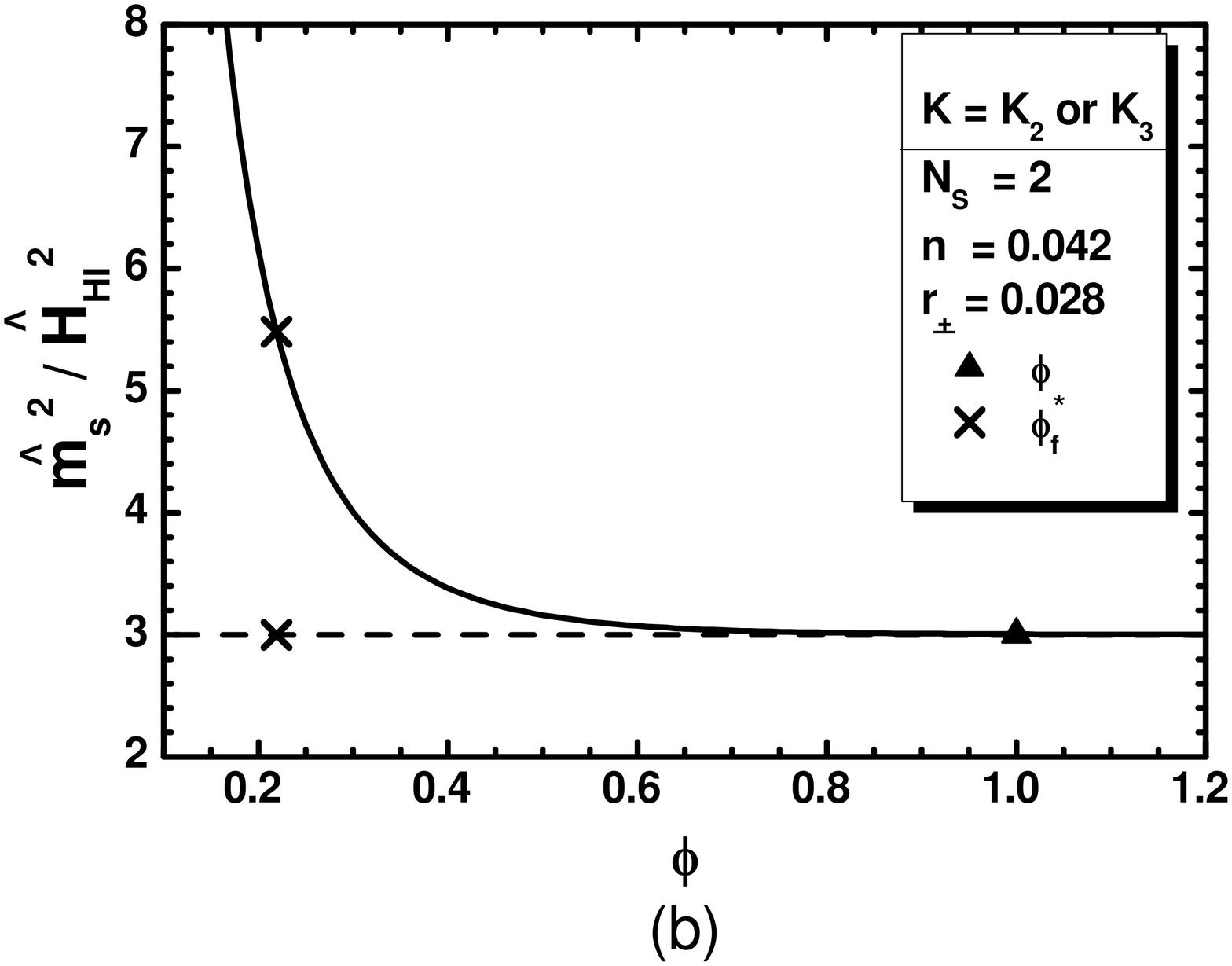,height=3.48in,angle=-90} \hfill
\end{minipage}
\hfill \caption[]{\sl \small The ratio $\what m^2_{s}/\Hhi^2$ as a
function of $\sg$  for $\sgx=1$ and $(n,\rs)=(0.042,0.028)$
computed by the exact numerical (solid line) or the approximate
analytic (dashed line) formulas. We set {\sffamily\ssz (a)}
$K=K_{1}$ and {\sffamily\ssz (b)} $K=K_{2}$ or $K_{3}$ with
$N_S=2$. The values corresponding to $\sgx$ and $\sgf$ are also
depicted.}\label{fig}
\end{figure}

To consolidate our inflationary setting we have to check the
stability of the trajectory in \Eref{infltr} with respect to the
fluctuations of the non-inflaton fields. Approximate expressions
for the relevant mass-squared spectrum are arranged in Table~3.
These expressions assist us to appreciate the role of $0<N<6$
[$0<N_S<6$] in retaining positive and heavy enough $\what m^2_{s}$
for $K=K_1$ [$K=K_2$ or $K_3$]. Indeed, $\what
m^2_{s}\gg\Hhi^2=\Vhi/3$ for $\sgf\leq\sg\leq\sgx$ -- where $\sgf$
is the value of $\sg$ at the end of HI -- as shown in
\sFref{fig}{a} [\sFref{fig}{b}] for $K=K_1$ [$K=K_2$ or $K_3$],
$\sgx=1$ (corresponding to $\cm=148$) and $(n,\rs)=(0.042,0.028)$
-- see below. From these plots we also infer that the approximate
formulas are quite precise for the largest part of the
inflationary period. In Table~3 we display also the mass,
$M_{BL}$, of the gauge boson $A_{BL}$ -- which signals the fact
that $U(1)_{B-L}$ is broken during HI -- and the masses of the
corresponding fermions. Inserting the derived mass spectrum in the
well-known Coleman-Weinberg formula, we can find the one-loop
radiative corrections, $\dV$ to $\Vhi$. It can be verified that
our results are immune from $\dV$, provided that the
renormalization group mass scale $\Lambda$, is determined
conveniently -- see \cref{jhep}.

\section{Results}\label{res}

The free parameters of our setting are $n, \rpm=\cp/\cm$ and
$\ld/\cm$ since if we perform the rescalings
$\Phi\to\Phi/\sqrt{\cm}~~~
\mbox{and}~~~\bar\Phi\to\bar\Phi/\sqrt{\cm}$ we see that $W$
depends on $\ld/\cm$ and $K$ on $n$ and $\rpm$. In \Sref{res2} we
confront the models with the observations and in \Sref{res1} we
show that these do not face any problem with the perturbative
unitarity.

\subsection{Testing Against Observations}\label{res2}

\begin{figure}[!t]\vspace*{-.12in}
\hspace*{-.19in}
\begin{minipage}{8in}
\epsfig{file=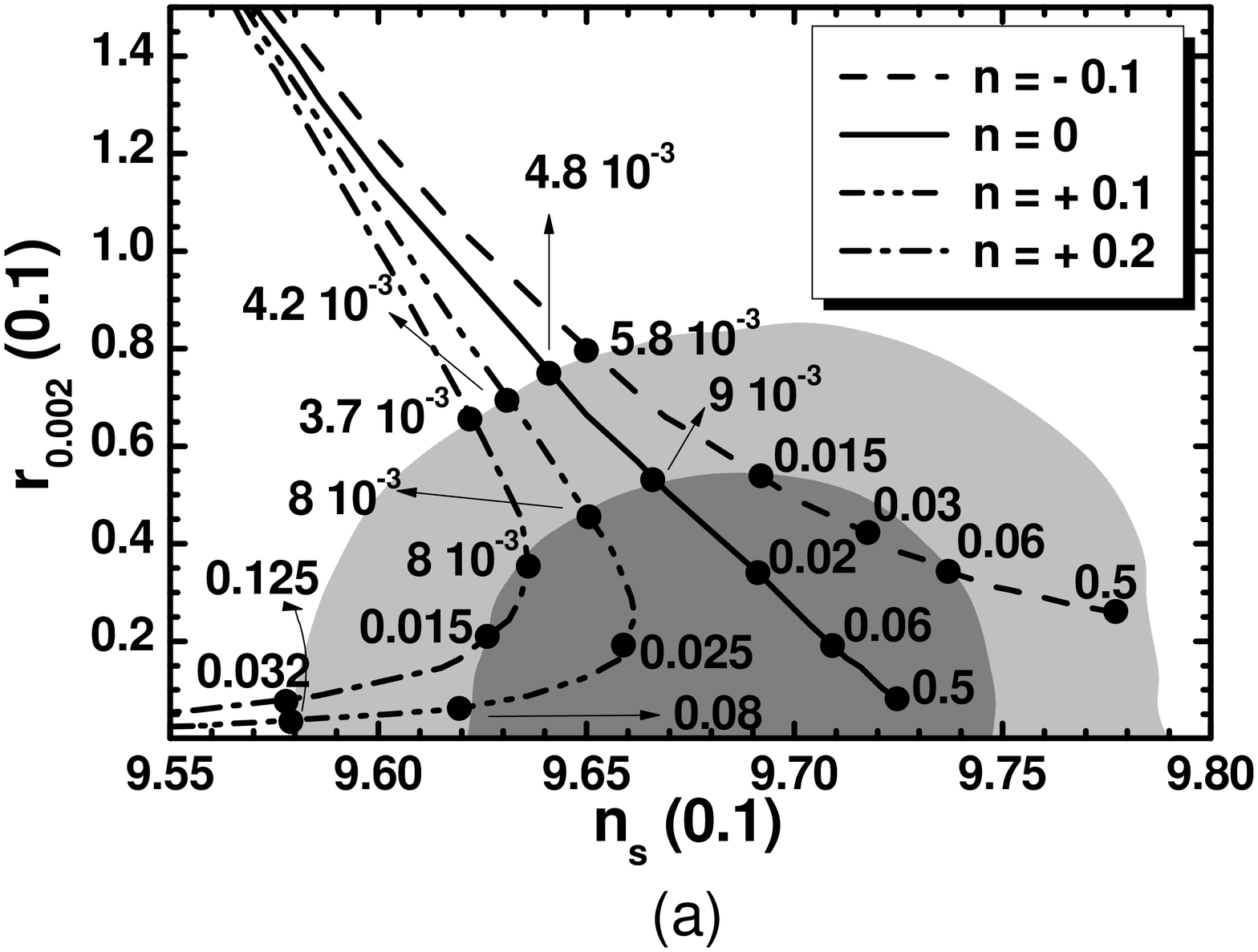,height=3.48in,angle=-90} \hspace*{-1.2cm}
\epsfig{file=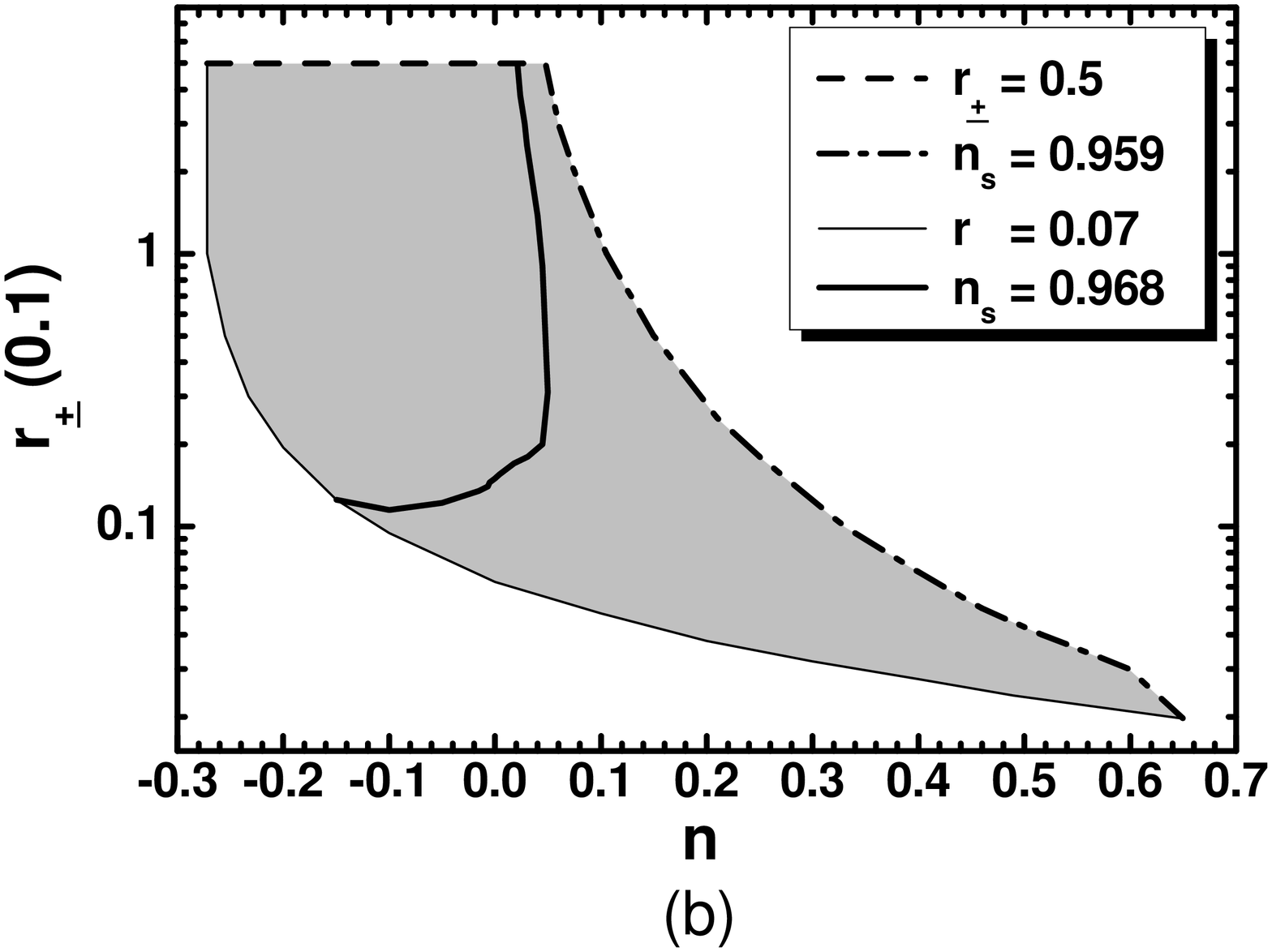,height=3.48in,angle=-90} \hfill
\end{minipage}
\hfill \caption{\sl\small {\sffamily\ftn (a)} Allowed curves in
the $\ns-\rw$ plane for $n=-0.1, 0, 0.1, 0.2$  with the $\rs$
values indicated on the curves -- the marginalized joint $68\%$
[$95\%$] regions from \plk, BAO and BK14 data are depicted by the
dark [light] shaded contours. {\sffamily\ftn (b)} Allowed (shaded)
regions in the $n-\rs$ plane. In both graphs we use $K=K_{2}$ or
$K_{3}$ with $N_S=2$. The conventions adopted for the various
lines are shown.}\label{fig1}
\end{figure}

To compare the predictions of our models with the observations, we
first compute -- applying standard formulas -- the number, $\Ns$,
of e-foldings that the scale $\ks$ experiences during HI and the
amplitude, $\As$, of the power spectrum of the curvature
perturbations generated by $\sg$ for $\sg=\sgx$. These observables
must be compatible with the requirements \cite{plin}, i.e.,
$\Ns\simeq58$ and $\sqrt{\As}\simeq4.627\cdot10^{-5}$ which assist
us in deriving $\sgx$ and $\ld/\cm$ as functions of $n$ and $\rs$.
We then extract the spectral index, $\ns$, its running, $\as$, and
the tensor-to-scalar ratio, $r$. These must be in agreement with
the fitting of the \plk, \emph{Baryon Acoustic Oscillations}
({\sf\ftn BAO}) and \bcp\ data ({\sf\ftn BK14}) \cite{plin,gwsnew}
with $\Lambda$CDM$+r$ model, depicted by gray and dark gray
contours in \sFref{fig1}{a}. The various lines represent the
theoretically allowed values for $K=K_2$ or $K_3$ and various
$n$'s as shown in the legend. The variation of $\rs$ is shown
along each line. For low enough $\rs$'s -- i.e. $\rs\leq0.0005$ --
the various lines converge to $(\ns,\rw)\simeq(0.947,0.28)$
obtained within quatric inflation defined for $\cp=0$. Increasing
$\rs$ the various lines enter the observationally allowed regions
and cover them allowing us to define a minimal and maximal $\rs$
corresponding to a maximal and minimal $r$ respectively. The lines
with $n>0$ [$n<0$] cover the left lower [right upper] corner of
the allowed range. In conclusion, the observationally favored
region can be wholly filled varying conveniently $n$ and $\rs$.

Varying continuously these parameters, we delineate the allowed
region of our models in \sFref{fig1}{b}. The conventions adopted
for the various boundaries are shown in the legend of the plot. In
particular, we take into account \cite{plin} the upper bound
$r\leq0.07$ and the lower bound $\ns\geq0.959$. Fixing $\ns$ to
its central value we obtain the thick solid line along which we
get clear predictions for $(n,\rs)$ and the remaining inflationary
observables. Namely, for $\ns=0.968$ and $\Ns\simeq58$, we find
\beq\label{reseq} -1.21\lesssim {n\over0.1}\lesssim0.215,\>\>\>
0.12\lesssim {\rs\over0.1}\lesssim5,\>\>\> 0.4\lesssim
{r\over0.01}\lesssim7\>\>\>\mbox{and}\>\>\>0.25\lesssim
10^5{\ld\over \cm}\lesssim2.6\,. \eeq

The utilized in \Fref{fig} values $(n,\rs)=(0.042,0.025)$ yields
the central values of observables, i.e., $(\ns,r)=(0.968,0.028)$.
Hilltop HI is attained for $0<n\leq0.0215$ and there, we get
$\Dex\gtrsim0.4$. The relevant tuning is therefore very mild. The
parameter $\as$ is confined in the range $-(5-6)\cdot10^{-4}$ and
so, our models are consistent with the fitting of data with the
$\Lambda$CDM+$r$ model \cite{plin}. Obviously, our models are
testable by the forthcoming experiments -- e.g., Core$+$,
LiteBird, Bicep3/{\it Keck Array} and SPIDER \cite{cmbol} --
searching for primordial gravity waves since $r\gtrsim0.0019$.

\renewcommand{\arraystretch}{1.}


\subsection{Perturbative Unitarity}\label{res1}

As can be seen numerically, there is a relatively large  lower
bound on $\cm$ for every $\rs$ above which $\sgx\leq1$. This fact
stabilizes our proposal against corrections from higher order
terms of the form $(\phc\phcb)^p$ with $p>1$ in $W$ -- see
\Eref{W}. Moreover, this fact does not jeopardize the validity of
the corresponding effective theory since these respect
perturbative unitarity up to $\mP=1$ as can be inferred by
analyzing the small-field behavior of our models. To this end, we
expand about $\vev{\phi}=M\ll1$ in terms of $\se$ the second term
in the right hand side of \Eref{Saction1} for $\mu=\nu=0$ and
$\Vhi$ in \Eref{Vhi}. Our results can be written as
\beq J^2
\dot\phi^2\simeq\lf1+3N\rs^2\what{\sg}^2-5N\rs^3\what{\sg}^4+\cdots\rg\dot\se^2\>\>\>\mbox{and}\>\>\>
\Vhi\simeq\frac{\ld^2\what{\sg}^4}{16\cm^{2}}\lf1-2(1+n)\rs\what{\sg}^{2}+\cdots\rg\,.\eeq
From the expressions above we conclude that our models are
unitarity safe up to $\mP$ since $\rs\leq1/N$ as shown below
\Eref{can}.

\section{Conclusions}\label{con}

We reviewed the implementation of kinetically modified non-minimal
HI in the context of SUGRA. The models are tied to the super-{}and
\Ka s given in Eqs.~(\ref{W}) and (\ref{K1}) -- (\ref{K3}).
Prominent in this setting is the role of a softly broken
shift-symmetry whose violation is parameterized by the quantity
$\rs=\cp/\cm$. Variation of $\rs$ in the range $(0.012-1/N)$
together with the variation of $n$ -- defined in \Eref{ndef} -- in
the range $(-0.121-0.0215)$ assists in fitting excellently the
present observational data and obtain $r$'s which may be tested in
the near future. These inflationary solutions can be attained even
with subplanckian values of the inflaton requiring large  $\cm$'s
and without causing any problem with the perturbative unitarity.


\def\ijmp#1#2#3{{\emph{Int. Jour. Mod. Phys.}}
{\bf #1},~#3~(#2)}
\def\plb#1#2#3{{\emph{Phys. Lett.  B }}{\bf #1},~#3~(#2)}
\def\zpc#1#2#3{{Z. Phys. C }{\bf #1},~#3~(#2)}
\def\prl#1#2#3{{\emph{Phys. Rev. Lett.} }
{\bf #1},~#3~(#2)}
\def\rmp#1#2#3{{Rev. Mod. Phys.}
{\bf #1},~#3~(#2)}
\def\prep#1#2#3{\emph{Phys. Rep. }{\bf #1},~#3~(#2)}
\def\prd#1#2#3{{\emph{Phys. Rev.  D }}{\bf #1},~#3~(#2)}
\def\npb#1#2#3{{\emph{Nucl. Phys.} }{\bf B#1},~#3~(#2)}
\def\npps#1#2#3{{Nucl. Phys. B (Proc. Sup.)}
{\bf #1},~#3~(#2)}
\def\mpl#1#2#3{{Mod. Phys. Lett.}
{\bf #1},~#3~(#2)}
\def\arnps#1#2#3{{Annu. Rev. Nucl. Part. Sci.}
{\bf #1},~#3~(#2)}
\def\sjnp#1#2#3{{Sov. J. Nucl. Phys.}
{\bf #1},~#3~(#2)}
\def\jetp#1#2#3{{JETP Lett. }{\bf #1},~#3~(#2)}
\def\app#1#2#3{{Acta Phys. Polon.}
{\bf #1},~#3~(#2)}
\def\rnc#1#2#3{{Riv. Nuovo Cim.}
{\bf #1},~#3~(#2)}
\def\ap#1#2#3{{Ann. Phys. }{\bf #1},~#3~(#2)}
\def\ptp#1#2#3{{Prog. Theor. Phys.}
{\bf #1},~#3~(#2)}
\def\apjl#1#2#3{{Astrophys. J. Lett.}
{\bf #1},~#3~(#2)}
\def\n#1#2#3{{Nature }{\bf #1},~#3~(#2)}
\def\apj#1#2#3{{Astrophys. J.}
{\bf #1},~#3~(#2)}
\def\anj#1#2#3{{Astron. J. }{\bf #1},~#3~(#2)}
\def\mnras#1#2#3{{MNRAS }{\bf #1},~#3~(#2)}
\def\grg#1#2#3{{Gen. Rel. Grav.}
{\bf #1},~#3~(#2)}
\def\s#1#2#3{{Science }{\bf #1},~#3~(#2)}
\def\baas#1#2#3{{Bull. Am. Astron. Soc.}
{\bf #1},~#3~(#2)}
\def\ibid#1#2#3{{\it ibid. }{\bf #1},~#3~(#2)}
\def\cpc#1#2#3{{Comput. Phys. Commun.}
{\bf #1},~#3~(#2)}
\def\astp#1#2#3{{Astropart. Phys.}
{\bf #1},~#3~(#2)}
\def\epjc#1#2#3{{Eur. Phys. J. C}
{\bf #1},~#3~(#2)}
\def\nima#1#2#3{{Nucl. Instrum. Meth. A}
{\bf #1},~#3~(#2)}
\def\jhep#1#2#3{{\emph{JHEP} }
{\bf #1},~#3~(#2)}
\def\jcap#1#2#3{{\emph{JCAP} }
{\bf #1},~#3~(#2)}
\def\jcapn#1#2#3#4{{\sl JCAP }{\bf #1}, no. #4, #3 (#2)}
\def\prdn#1#2#3#4{{\sl Phys. Rev. D }{\bf #1}, no. #4, #3 (#2)}
\newcommand{\arxiv}[1]{{\ftn\tt  arXiv:#1}}
\newcommand{\hepph}[1]{{\ftn\tt  hep-ph/#1}}

\end{document}